\documentclass[runningheads]{llncs}

\usepackage{silence}
\usepackage[T1]{fontenc}
\usepackage{graphicx}
\usepackage{amsmath, amssymb}
\usepackage{narsese-symbol}
\usepackage{booktabs}
\spnewtheorem{define}{Definition}{\bfseries}{\rmfamily}

\begin{document}

\title{Anti-Goal Reasoning: Rethinking the Theory of Goal Reasoning in Non-Axiomatic Logic}
\titlerunning{Anti-Goal Reasoning}

\author{
Bowen Xu\inst{}\orcidID{0000-0002-9475-9434}
}

\authorrunning{B. Xu}

\institute{
Department of Computer and Information Sciences, Temple University, USA \\
\email{bowenxu.agi@gmail.com} \\
\vspace{0.3cm}
July 8, 2026
}

\maketitle

\begin{abstract}
Goal reasoning in Non-Axiomatic Logic (NAL) explains how an adaptive system
derives means for realizing desired events under insufficient knowledge and
resources. However, the representation of avoidance is less clear. A common
convention is to express ``avoid $G$'' as the goal sentence ``$\neg G!$'', but
this notation conflates two different readings: pursuing the negated event
$\neg G$, and avoiding the positive event $G$. This paper shows that the
conflation can produce a paradoxical case in which an avoidance
intention is converted into a positive goal to act merely because acting is
usually followed by the absence of hurt. Starting from NAL's basic definition of goals, the framework is extended
with a corresponding definition of anti-goals, so that avoidance can be
represented without treating it as the pursuit of a negated event. Finally, a
mental operation, $\op{prevent}$, is introduced to connect anti-goal reasoning
with ordinary goal reasoning in cases of active prevention. Four minimal case studies
check that the resulting rules distinguish pursuit, passive avoidance, active
prevention, and withholding action to preserve a desired event.
\end{abstract}

\section{Introduction}

Realizing beneficial events and avoiding harmful events are both basic
functions of an adaptive system.

In Non-Axiomatic Logic (NAL)~\cite{wang2025nal-ed2}, a logic for constructing
adaptive systems under insufficient knowledge and resources, these functions are
handled through \textit{goal reasoning}. A reasoning system based on NAL is
called a Non-Axiomatic Reasoning System (NARS). A \textit{goal} in NARS is an
event the system desires to realize. A typical backward goal inference has the
following form: if a result usually follows from a condition, then, to realize
the result, the system should try to realize the condition.

Two issues remain unclear in the current account of goal reasoning in NAL.
\begin{enumerate}
    \item Reasoning about avoidance has not been specified. In particular, how
    should ``avoid an event'' be represented? A common convention is to use a
    negated goal form, but this convention leads to the problem developed in
    Sec.~\ref{sec:paradox}.
    \item The desire-value functions for concrete goal-reasoning rules may not have been
    well justified, especially in existing implementations of NARS, such as OpenNARS~\cite{hammer2016opennars} and ONA~\cite{hammer2020ona}.
\end{enumerate}

This paper presents a simple NAL goal-reasoning example that produces a
paradoxical result. I argue that the origin of the paradox is the semantic ambiguity
of ``goal negation''. To handle avoidance more precisely, the paper re-examines the
definitions of goal and desire-value. It then introduces the opposite of the desired state as the state to be avoided, defines a separate anti-goal form, and clarifies how desire-value functions for goal and anti-goal rules should be chosen. As a theoretical supplement, the mental operation $\op{prevent}$ is used to connect goal reasoning with anti-goal reasoning in cases where active prevention is needed.

\section{Overview of Non-Axiomatic Logic}
\label{sec:nal-overview}

Before turning to the core issue, this section briefly reviews the parts of NAL necessary for
this paper. The overview is not meant to be complete. It only covers the
concepts needed to understand the paradox in Sec.~\ref{sec:paradox}, the
analysis in Sec.~\ref{sec:analysis}, and the proposed extension in
Sec.~\ref{sec:potential-solution}, without requiring the reader to consult the
full NAL book~\cite{wang2025nal-ed2}.

\subsection{Inheritance Logic}
The formal definitions of NAL are obtained by first considering an idealized
situation and then adjusting it to the Assumption of Insufficient Knowledge and
Resources (AIKR)~\cite{wang2025nal-ed2}. The idealized situation is described
by Inheritance Logic (IL), where the system is assumed to have sufficient
knowledge and resources and uses the Closed-World Assumption: Its ideal
experience is a finite set $K$ of statements, and its knowledge $K^*$ is the
transitive closure of $K$ under the relevant inference rules. In this setting,
truth is binary: a statement is true when it belongs to $K^*$, otherwise it is false.

NAL is the AIKR reinterpretation of this idealized situation, not a separate
execution of IL inside NARS. Under AIKR, inputs arrive over time and
computational resources are limited, so object-level knowledge cannot be treated
as permanent axioms. Instead, knowledge items are revisable, binary truth becomes
evidential support, and truth-preserving inference becomes inference with
truth-value functions.

\subsection{Implication and Negation in NAL}

IL defines an \textit{implication} statement ``$A \imply C$'' as true \textit{iff}
statement $A$ derives statement $C$. Statement $A$ is then included in the
\textit{sufficient conditions} set of $C$ (denoted as $C^S$), while statement
$C$ is included in the \textit{necessary conditions} set of $A$ (denoted as
$A^N$). This semantics makes the amount of evidence countable.

To extend IL to NAL, the amount of positive and negative evidence for
``$A \imply C$'' is defined. The positive evidence (denoted by $w^+$) is included
in $(A^S \cap C^S)$ and $(A^N \cap C^N)$, while the negative evidence (denoted
by $w^-$) is included in $(A^S - C^S)$ and $(C^N - A^N)$.

By counting the number of elements in the sets, frequency and confidence are defined,
\begin{equation*}
f = \frac{w^+}{w},\quad c = \frac{w}{w+k},
\end{equation*}
where the total evidence is $w=w^+ + w^-$, and $k$ is a constant (usually $k=1$). \textit{Truth-value} is defined as a pair of frequency and confidence
\begin{equation*}
\truthv{f}{c}.
\end{equation*}
In NAL, the truth-value of a knowledge item is either supplied by input or
derived from existing knowledge. It is not calculated by directly counting the
ideal evidence sets. Instead, the evidence semantics provides the unit of
measure, much as a physical unit is defined by a standard without requiring each
measurement to be compared directly with that standard.

In IL, the ideal experience contains only positive knowledge, while negative
knowledge is represented implicitly as the complement of the former: statements
outside the system's knowledge are assumed false. When IL is extended into NAL,
the negation operator is defined as swapping the positive and negative evidence
for a statement. In NAL, knowledge outside the system's memory is
\textit{unknown} (with confidence $c=0$) rather than false, while a totally
false statement has frequency $f=0$ and confidence $c > 0$.

According to the evidential semantics,
``$A \imply C$'' is supported by the conjunction of $A$ and $C$, ``$(A \wedge C)$'',
and opposed by the conjunction of $A$ and $\neg C$, ``$(A \wedge \neg C)$''.
Therefore, ``$\neg (A \imply C )$'' is equivalent to ``$A \imply (\neg C)$''.
Statement ``$\neg (A \imply C )$'' means the exact opposite of ``$C$ can be derived
from $A$'', namely, ``$\neg C$ can be derived from $A$''.

\subsection{Goals and Backward Goal Inference}

In NAL, an \textit{event} is a statement with temporal attributes. Its
truth-value is time-dependent, in the sense that the evidential support
summarized in the truth-value is valid only within a certain duration.

A goal can be understood as an event that the system desires to realize, and
the strength of this desire is called its desire-value. In notation, a
goal with event content $G$ is written as ``$G!$''. Definitions~\ref{def:goal}
and~\ref{def:desire-value} restate the corresponding definitions from NAL.

NAL specifies the principles of backward inference by ``meta-rules''. In
schematic form, the meta-rule for goal reasoning is
\begin{equation}
    \text{IF} ~\text{``}\{J,\ J_{G'}\}\vdash J_G\text{''}, \quad
    \text{THEN}~ \text{``}J,\ G!\vdash G'!\text{''},
    \label{eq:analysis-backward-meta-rule}
\end{equation}
where $J_G$ and $J_{G'}$ are judgments whose contents are $G$ and $G'$,
respectively. For a unary judgment transformation, the corresponding form is
\begin{equation}
    \text{IF} ~\text{``}J_{G'}\vdash J_G\text{''}, \quad
    \text{THEN}~ \text{``}G!\vdash G'!\text{''}.
    \label{eq:analysis-backward-meta-rule-unary}
\end{equation}
For example, a forward deduction rule in NAL is
``$\{A \imply C ~\truthv{f_1}{c_1},~ A~\truthv{f_2}{c_2}\}\vdash C ~\truthf{ded}$''\footnote{Here $F_{\mathrm{ded}}$ is the standard NAL deduction truth-value function: for premises with truth-values $\truthv{f_1}{c_1}$ and $\truthv{f_2}{c_2}$, the conclusion receives $\truthv{f_1 f_2}{f_1 f_2 c_1 c_2}$.}: if $A$ implies $C$ and $A$ is
true, then $C$ is true. The corresponding backward goal-inference rule is
``$\{A \imply C,~ C!\}\vdash A!$'': if $A$ implies $C$, then $A$
can be treated as a means for realizing the goal $C$. 
The desire-value of the conclusion is calculated from the truth-value of one premise and the desire-value of the other. This mapping is called a \textit{desire-value function}.
A desire-value function is not itself a truth-value function for two judgments, but it may either use the same calculation formula as the corresponding truth-value function or be specifically designed, with notation such as $\truthf{ded}$ and, for example, $\truthf{strong}$\footnote{Here $\truthf{strong}$ is not a function defined in NAL~\cite{wang2025nal-ed2}; it is only an example of a symbol that a designer may introduce.}.

However, the desire-value functions for concrete goal rules are not explicitly
specified in \cite{wang2025nal-ed2}. In some previous practice, the desire-value function for the deduction
rule ``$\{A\imply C,~A!\}\vdash C!$'' is $\truthv{f_1f_2}{c_1c_2}$, and that
for the negation rule ``$G!\vdash\neg G!$'' is $\truthv{1-f}{c}$.


\section{Paradox}
\label{sec:paradox}

The problem can be exposed by combining three assumptions\footnote{These assumptions are not those taken in NAL except \eqref{eq:A1}. Assumption~\eqref{eq:A2} is a self-evident belief, while \eqref{eq:A3} is a convention usually adopted in previous NARS research.} that are each
plausible in isolation. The first two assumptions concern ordinary backward
motivational reasoning: a sufficient means to a desired consequence should
inherit goal support, while a sufficient cause of an avoided consequence should
itself be avoided:
\begin{align}
\mathrm{\parbox[t]{0.8\linewidth}{Given judgment ``$A \imply C$'', if the system wants to \textit{realize} $C$, it should be able to derive through strong inference that it should \textit{realize} $A$.}}\tag{A1}\label{eq:A1}\\
\mathrm{\parbox[t]{0.8\linewidth}{Given judgment ``$A \imply C$'', if the system wants to \textit{avoid} $C$, it should be able to derive through strong inference that it should \textit{avoid} $A$.}}\tag{A2}\label{eq:A2}
\end{align}

The third assumption is an implicit notational convention often used in
previous NARS practice, even though it is seldomly stated explicitly: avoidance is
represented by putting negation inside an ordinary goal sentence. That is,
\begin{equation}
\text{``avoid $G$'' is defined as ``$\neg G!$''.} \tag{A3}\label{eq:A3}
\end{equation}

Now consider the following light-press case:
\begin{align}
    (light, \op{press}) \predimply hurt ~\truthv{0.7}{0.99},\tag{P1}\label{eq:P1}\\
    \op{press} \predimply \neg{hurt}  ~\truthv{0.9}{0.99},\tag{P2}\label{eq:P2}\\
    \neg hurt!  ~\truthv{1.0}{0.99}.\tag{G1}\label{eq:G1}
\end{align}
The intended reading is:
\begin{itemize}
    \item when the light signal is present, pressing the button is
    \textit{often} followed by $hurt$;
    \item pressing the button in general is \textit{usually} followed by
    a negative observation of hurt;\footnote{In NARS, a negative observation, such as $\neg hurt ~\truthv{0.0}{0.5}$, is usually produced by a failed anticipation.}
    \item the system wants to avoid $hurt$, represented as ``$\neg hurt!$''
    according to assumption~\eqref{eq:A3}.
\end{itemize}
Intuitively, this case suggests the following practical expectation:
\begin{equation}
\mathrm{\parbox[t]{0.8\linewidth}{A rational subject will not press the button when a light signal occurs, and there is no need to press the button when no light occurs.}}\tag{B1}\label{eq:B1}
\end{equation}

However, when no light is present, \eqref{eq:P2} and \eqref{eq:G1} support
\begin{equation}
    \op{press}!, \tag{G2}\label{eq:G2}
\end{equation}
by \eqref{eq:A1}: if pressing usually leads to $\neg hurt$, and ``$\neg hurt!$''
is treated as an ordinary goal, then pressing appears to be a means to that
goal. On the surface, this derivation also matches the form of the meta-rule in
\eqref{eq:analysis-backward-meta-rule}. This is already unintuitive. The fact
that pressing is followed by non-hurt does not show that pressing is needed to
avoid hurt.

The conflict becomes sharper when the light signal occurs:
\begin{equation}
    light. ~\truthv{1.0}{0.99}. \tag{P3}\label{eq:P3}
\end{equation}
In this context, the ordinary belief \eqref{eq:P1} says that pressing under the
light is a likely cause of $hurt$. Since \eqref{eq:G1} is intended to express
avoidance of $hurt$, this belief supports the practical expectation stated in
\eqref{eq:B1}: the system should not press under the light.

However, \eqref{eq:G2} remains derivable. Moreover, since \eqref{eq:P2} has
a higher truth-value frequency than \eqref{eq:P1} (\textit{i.e.},$0.9>0.7$), the desire-value
propagated from \eqref{eq:P2} to \eqref{eq:G2} should also be stronger than
the desire-value propagated from \eqref{eq:P1} to ``$\neg\op{press}!$''.
Therefore, the system may press even in the light condition. The resulting
behavior can be:
\begin{equation}
\mathrm{\parbox[t]{0.8\linewidth}{The system presses the button when no light occurs, and it still presses the button when a light signal occurs.}}\tag{C1}\label{eq:C1}
\end{equation}
As a result, the system presses in order to avoid hurt, but in the light condition pressing
is precisely what the ordinary belief \eqref{eq:P1} says tends to lead to hurt.
This is the intended paradox: the premises produce a positive goal to press,
while the normal practical belief \eqref{eq:B1} says that pressing is not
needed in general and should be withheld under the light.

In short, assumptions \eqref{eq:A1}--\eqref{eq:A3}, together with the intuitive
requirement \eqref{eq:B1}, generate the paradoxical consequence
\eqref{eq:C1} conflicting to \eqref{eq:B1}.

\section{Analysis}
\label{sec:analysis}

Why does the paradox occur? I argue that it stems from the semantic ambiguity
of goal negation in
assumption~\eqref{eq:A3}. The same expression, ``$\neg G!$'', is mistakenly used for two different readings: a goal whose content is $\neg G$, and an anti-goal whose avoided event is $G$. These two readings have different evidential meanings in backward inference.

The desire-value of a goal ``$G!~\truthv{f}{c}$'' is defined as the truth-value of
\begin{equation}
    G\imply\tilde{D}~\truthv{f}{c},
    \label{eq:analysis-goal-semantics}
\end{equation}
where $\tilde{D}$ is a virtual term, defined at the meta-level, that summarizes
the overall desired state of the system.
The negation of this goal statement has the form
\begin{equation}\label{eq:analysis-undesired}
\neg(G \imply \tilde{D}),
\end{equation}
which is the exact opposite of ``$G \imply \tilde{D}$'' by swapping the
positive and negative evidence, according to NAL~\cite{wang2025nal-ed2}.
However, the expression ``$\neg(G \imply \tilde{D})$'' can be confused with
two different readings:
\begin{align}
    (\neg G)\imply\tilde{D},
    \label{eq:analysis-neg-event}\tag{I1}\\
    G\imply(\neg\tilde{D}).
    \label{eq:analysis-neg-desired-state}\tag{I2}
\end{align}
These readings are not equivalent. In \eqref{eq:analysis-neg-event}, negation
is applied to the event $G$; in \eqref{eq:analysis-neg-desired-state},
negation is applied to the virtual desired-state term $\tilde{D}$. The two
placements therefore have different consequences in backward reasoning.

\subsection{Pursuit of the negated event}
If negation applies to the event inside the goal, then
\begin{equation}
    (\neg G)!
\end{equation}
means
\begin{equation}
    (\neg G)\imply\tilde{D}.
    \label{eq:analysis-pursue-negation}
\end{equation}
This means that realizing the event $\neg G$ contributes to the
desired state. It therefore represents pursuit of $\neg G$.

Consequently, an event that leads to $\neg G$ can inherit goal support. For
example,
\begin{equation}
\begin{aligned}
    X\Rightarrow\neg G,
    \qquad
    (\neg G)\imply\tilde{D}
    \quad\vdash\quad
    X\imply\tilde{D}.
\end{aligned}
\label{eq:analysis-pursue-negation-backward}
\end{equation}
Thus, ``$(\neg G)!$'' can provide a reason for pursuing any event predicted to
produce $\neg G$. This inference is appropriate when $\neg G$ is genuinely an
event that the system wants to realize. It does not follow merely from having
reasons against pursuing $G$, and it does not by itself mean that $G$ is to be
prevented.

\subsection{$G$ leading to an undesired state}

Interpretation \eqref{eq:analysis-neg-desired-state}
may initially appear closest to the intuitive meaning of avoidance. It seems to
say that $G$ leads to an ``undesired state.''

Suppose, provisionally, that $\neg\tilde{D}$ is interpreted as the ordinary
complement of $\tilde{D}$. On this binary reading, any occurrence of $G$ that
is not followed by the desired state would count as an occurrence of
$G$ followed by the undesired state summary. Equation
\eqref{eq:analysis-undesired} would then express the exact evidential opposite
of ``$G\imply\tilde{D}$''.

However, Eq.~\eqref{eq:analysis-neg-desired-state} faces a fundamental formal
problem. The term $\tilde{D}$ is not an object-level term; it is a virtual term
introduced at the meta-level to summarize the system's desired state. In the
current account of NAL, this virtual term does not have a defined negation
semantics. Consequently, ``$\neg\tilde{D}$'' is presently undefined. The phrase
``undesired state'' gives it an intuitive gloss, but it does not establish a
formal NAL meaning. Therefore, Eq.~\eqref{eq:analysis-neg-desired-state} is
formally unspecified.

\subsection{Source of the paradox}

The source of the paradox is a change of reading for the same sentence. In the
derivation of~\eqref{eq:G2}, ``$\neg hurt!$'' is read according to
Eq.~\eqref{eq:analysis-pursue-negation}, namely as pursuit of the negated event
``$(\neg hurt)\imply\tilde{D}$''. Together with~\eqref{eq:P2},
``$\op{press}\predimply\neg hurt$'', ordinary backward goal reasoning derives
``$\op{press}!$''. Under this reading, the inference is internally coherent:
pressing is treated as a means to realize $\neg hurt$.

Under the intended avoidance reading of assumption~\eqref{eq:A3}, however,
``$\neg hurt!$'' should mean that $hurt$ is to be avoided, not that $\neg hurt$
is to be pursued. In the light condition, this reading must be combined
with~\eqref{eq:P1}, where pressing under the light is a likely cause of
$hurt$. The practical expectation is therefore \eqref{eq:B1}, not the behavior
summarized in~\eqref{eq:C1}.

In a nutshell, the root cause of the paradox is the conflation of two readings of the same
notation. 

The conflict between~\eqref{eq:C1} and~\eqref{eq:B1} shows that 
the negative form of goal, the reasoning of avoidence, and the meta-rules for goal reasoning need further clarification.

\section{Potential Solution}
\label{sec:potential-solution}

The proposed solution is to separate two representations that were conflated in
assumption~\eqref{eq:A3}: a goal whose event content is $\neg G$, and an
anti-goal whose avoided event is $G$. The rest of this section follows the
construction methodology used in~\cite{wang2025nal-ed2}: start from an IL
idealization, then reinterpret it under NAL's evidential semantics. First, the
existing account of goals is recalled through the virtual desired-state term
$\tilde{D}$. Next, the interpretation of the desired-state and undesired-state
summaries in IL is made explicit. The binary IL distinction is then quantified as
desire-value and aversion-value in NAL. Finally, the
corresponding inference rules are designed according to the definitions and theorems.

The definition of \textit{goal} remains the same as in \cite{wang2025nal-ed2}
\begin{define}
\label{def:goal}
A \textit{goal} ``$G!$'' is a sentence containing an event that the system
desires to realize.
\end{define}
But the virtual term $\tilde{D}$ is made explicit in IL:
\begin{define}
\label{def:virtual-D}
In IL, a goal implies the desired-state $\tilde{D}$, where $\tilde{D}$ is a
virtual statement summarizing the system's current goals. 
\end{define}
Here $\tilde{D}$ is ``virtual'' in the sense that it is not a concrete
statement in memory, but a meta-level conceptual construction used in the design
of the system. It is introduced only as a summary term in the definition of
goals, and it has no further \textit{necessary conditions}; that is,
\begin{define} The nececesy conditions of $\tilde{D}$ only contains $\tilde{D}$ itself.
$$\{x \mid \tilde{D} \imply x\} = \{\tilde{D}\} .$$
\end{define}
That is, any statement in $K^*$ is not in the necessary conditions of $\tilde{D}$:
\begin{theorem}$\{x \mid \tilde{D} \imply x \wedge x \in K^*\} = \varnothing .$
\end{theorem}
As a variant of Theorem 9.2 in \cite{wang2025nal-ed2}, it is evident that
\begin{theorem}
In IL, if $G$ is a desired statement in $K^*$, then
$$(G \imply \tilde{D}) \equiv (G^S \subseteq \tilde{D}^S).$$
\end{theorem}

We call a statement, that is undesired, aversive, or avoided by the system, as ``anti-goal''.
\begin{define}
\label{def:goal}
An \textit{anti-goal} ``$G\mbox{!`}$'' is a sentence containing an event that the system avoid to realize.
\end{define}
According to IL, any unknown knowledge are assumed false. With the similar principle, any statement that does not imply $\tilde{D}$ is avoided.
\begin{define}
In IL, if $G$ is not a desired statement in $K^*$, then $G$ is an anti-goal.
\end{define}

To extend goal-related representation and inference from IL to NAL, the binary representation should be converted to quantified representation. 
To bridge the gap, the evidential support of a goal is specified:
\begin{theorem}
For a goal ``$G!$'', its evidence includes statements in $G^S$. Among them,
statements in $(G^S \cap \tilde{D}^S)$ are positive evidence, while
statements in $(G^S - \tilde{D}^S)$ are negative evidence.
\end{theorem}
This theorem follows directly from Definition 9.4 of \cite{wang2025nal-ed2}.
\textit{Desire-value} then quantifies the extent of wanting to realize an event in NAL:
\begin{define}
\label{def:desire-value}
The \textit{desire-value} of an event measures the extent to which a desired
state is implied by the event. The desire-value of event $S$ is the truth-value
of the implication statement ``$S \imply \tilde{D}$'', where $\tilde{D}$ is a
virtual statement summarizing the system's current goals.
\end{define}
This definition restates Definition 12.4 in \cite{wang2025nal-ed2}.
In IL, the sufficient conditions of
$\tilde{D}$ include the desired statements, while statements outside them are
considered undesired in the relevant sense. When this account is extended to
NAL, a virtual statement $\neg\tilde{D}$ to represent the opposite
of the desired-state in NAL.
\begin{define}\label{def:aversion-value}
The \textit{aversion-value} of an event measures the extent to which the
opposite of the desired state is implied by the event. The aversion-value of
event $S$ is the truth-value of the implication statement
``$S \imply \neg\tilde{D}$'', where $\neg\tilde{D}$ is the virtual statement for
the opposite of the desired state:
$$
\neg (G \imply \tilde{D}) \equiv G \imply (\neg\tilde{D}).
$$
\end{define}
Some events are neither desired nor undesired, meaning the system has weak or even no motivational attitude toward them. In the extreme case, both supporting and opposing evidence are $0$, which is reflected in the confidence value $c=0$.

According to Definitions~\eqref{def:desire-value} and \eqref{def:aversion-value},
for a goal, positive evidence measures support for realizing the event, whereas negative evidence measures opposition to realizing it.
Thus, a goal and an anti-goal are opposites of each other, and their evidence can be obtained from each other by swapping the positive and negative evidence. That is
\begin{theorem}
To avoid an event means not to realize it:
\[
G\textup{\mbox{!`}} \equiv G\imply(\neg\tilde{D})
\equiv \neg (G \imply \tilde{D}) \equiv \neg{G}!.
\]
\end{theorem}

Here the expression ``$\neg{G}!$'' is read at the level of the goal statement
``$G \imply \tilde{D}$''. It is not the ambiguous convention in \eqref{eq:A3},
where ``avoid $G$'' was identified with the ordinary goal whose event content is
$\neg G$. Symmetrically,

\begin{theorem}
To realize an event means not to avoid it:
\[
G! \equiv G\imply\tilde{D}
\equiv \neg(\neg(G\imply\tilde{D}))
\equiv \neg{G}\textup{\mbox{!`}}.
\]
\end{theorem}

These equivalences justify the negation rules between goal and anti-goal sentences:
\begin{equation}
\begin{aligned}
G! \vdash G\mbox{!`}~\truthf{neg}, \\
G\mbox{!`} \vdash G!~\truthf{neg}.
\end{aligned}
\end{equation}

Other desire-value functions for goal-inference rules can be obtained by the following process: after converting a goal or anti-goal into the form ``$G \imply \tilde{D}$'' or ``$G \imply \neg\tilde{D}$'' respectively, if there exists a corresponding forward inference rule, the backward rule adopts the same calculation formula as that rule's truth-value function. In this sense, $\truthf{ded}$ in a backward motivational rule names a desire-value function whose formula is copied from the corresponding truth-value function, rather than a truth-value function directly applied to a judgment. For example, given premises ``$A \imply C$'' and ``$C!$'', the latter can be transformed to ``$C\imply\tilde{D}$'', since the deduction rule ``$\{M\imply P,S\imply M\} \vdash S\imply P~\truthf{ded}$'' exists, the backward rule uses the corresponding desire-value function with the same formula as $\truthf{ded}$. The same process works with an anti-goal. Thereby, the following two rules are produced:
\begin{align}
\{A \imply C ~\truthv{f_1}{c_1},~ C!~\truthv{f_2}{c_2}\}
&\vdash A! ~\truthf{ded},\label{eq:ded-goal} \\
\{A \imply C ~\truthv{f_1}{c_1},~ C\mbox{!`}~\truthv{f_2}{c_2}\}
&\vdash A\mbox{!`} ~\truthf{ded}. \label{eq:ded-anti}
\end{align}
The form of these backward rules can still be understood as an instance of the
meta-rule in Eq.~\eqref{eq:analysis-backward-meta-rule}, but what requires separate
justification is the choice of desire-value function.
Generally speaking, if a goal-inference rule is produced, its symmetric form for anti-goal inference can be obtained at the same time.

For compound terms, decomposition rules allow a desired conjunction to propagate
goal support to its components. For example,
\begin{equation}
\{A~\truthv{f_1}{c_1}, ~(A \wedge C)!~\truthv{f_2}{c_2}\}
\vdash C! ~\truthf{ded}.
\label{eq:ded-seq-tail}
\end{equation}
Here, the desire-value function is not copied from $\truthf{int}$, which is used in the forward rule, ``$\{A, C\}\vdash (A\wedge C)~\truthf{int}$''~\cite{wang2025nal-ed2}. In contrast, the desire-value function stems from the following rule, ``$\{(A\wedge C)\imply D, A\}\vdash C\imply D~\truthf{ded}$''.

The design of temporal variants of backward rules, such as the rules that involves predictive implication (\textit{e.g.}, ``$A\predimply C$'') or sequential compound (\textit{e.g.}, ``$(A, C)$''), follows the same principles as designed in Chapter 11 of \cite{wang2025nal-ed2}.

In addition to the intrinsic evidential relation between goals and anti-goals, some situations require the system to derive a goal from an anti-goal, or vice versa. Such conversion can be driven by mental operation ``$\op{prevent}$'' -- if $G$ is an anti-goal, then preventing $G$ can become a goal. If $G$ is a goal, then preventing $G$ can become an anti-goal:
\begin{equation}\label{eq:prevent-activation}
\begin{aligned}
\op{prevent}(G)! &\equiv G\mbox{!`},\\
\op{prevent}(G)\mbox{!`} &\equiv G!.
\end{aligned}
\end{equation}
This is crucial for causal discovery. For example, under a high-temperature
condition, turning on a fan may be followed by no overheating, while not turning
on the fan is followed by overheating. The contrast supports the hypothesis that
turning on the fan prevents overheating in that condition.
This prevention is discovered by a contrastive procedure.

Rather than introducing new inference rules with global effects, the paper represents
this contrastive inference as a local object-level knowledge item:
\begin{equation}\label{eq:prevent-schema}
\begin{aligned}
&(((\$C \wedge \$A) \imply \neg \$R) \\
&\wedge \\
&((\$C \wedge \neg \$A) \imply \$R)) \\
&\imply \\
&((\$C \wedge \$A) \imply \op{prevent}(\$R)) ~\truthv{1.0}{0.99},
\end{aligned}
\end{equation}
where ``$\$$'' followed by a variable name indicates an \textit{independent
variable} in NAL~\cite{wang2025nal-ed2}; such a variable can be replaced by a
constant term. The knowledge item in \eqref{eq:prevent-schema} is attached to the ``$\op{prevent}$'' operation and is triggered conditionally.\footnote{A previous account of \textit{causal inference} in NARS~\cite{xu2024causal} did not address prevention. The present extension therefore enriches the causal-inference theory of NARS.}

\section{Case Studies}
\label{sec:case-studies}

This section gives a compact empirical check of the proposed distinction
between goals and anti-goals. The cases are deliberately minimal: they are constructed from a shared small vocabulary, including the sensory terms $light$, $food$, $hurt$, and the mental operation $\op{press}$. The point is not to evaluate a full-scale NARS system, but to verify
that the proposed semantics derives the expected operational motivation in the four situations required by the theory:
\begin{enumerate}
    \item act to realize a desired event;
    \item withhold an operation that would realize an undesired event;
    \item act to prevent an undesired event;
    \item withhold an operation that would prevent a desired event.
\end{enumerate}

A minimal system is implemented to run the cases. Each run checks whether the reasoning process derives the expected
operation-level item: a goal ``$\op{press}!$'' supports pressing, whereas an
anti-goal ``$\op{press}\mbox{!`}$'' suppresses pressing. Each run lasts 160
working cycles. During cycles 1--80, motor babbling is enabled so that the
system samples both pressing and not pressing. During cycles
81--160, motor output is driven by the operational goal inferred by the system.
To evaluate the system's overall behavioral performance, the simulated body is
equipped with fixed evaluative channels: each occurrence of $food$ contributes
$+1$ to cumulative reward, while each occurrence of $hurt$ contributes $-1$. The curves
therefore show the cumulative reward generated by these external channels; goals
and anti-goals are not defined from this scalar quantity.

Table~\ref{tab:minimal-light-press} summarizes the reasoning and behavioral
results of the four cases.

{
\setlength{\tabcolsep}{4pt}
\begin{table}
\centering
\caption{Reasoning and behavioral results of the four light-press cases.}
\label{tab:minimal-light-press}
\small
\begin{tabular}{p{0.28\linewidth}p{0.18\linewidth}p{0.22\linewidth}p{0.16\linewidth}}
\toprule
Case name & Input & Derived & Decision \\
\midrule
Do to Realize & ``$food!$'' & ``$\op{press}!$'' & press \\
Not Do to Avoid & ``$hurt\mbox{!`}$'' & ``$\op{press}\mbox{!`}$'' & no operation \\
Do to Avoid & ``$hurt\mbox{!`}$'' & ``$\op{press}!$'' & press \\
Not Do to Realize & ``$food!$'' & ``$\op{press}\mbox{!`}$'' & no operation \\
\bottomrule
\end{tabular}
\end{table}
}

\begin{figure}
\centering
\includegraphics[width=\linewidth]{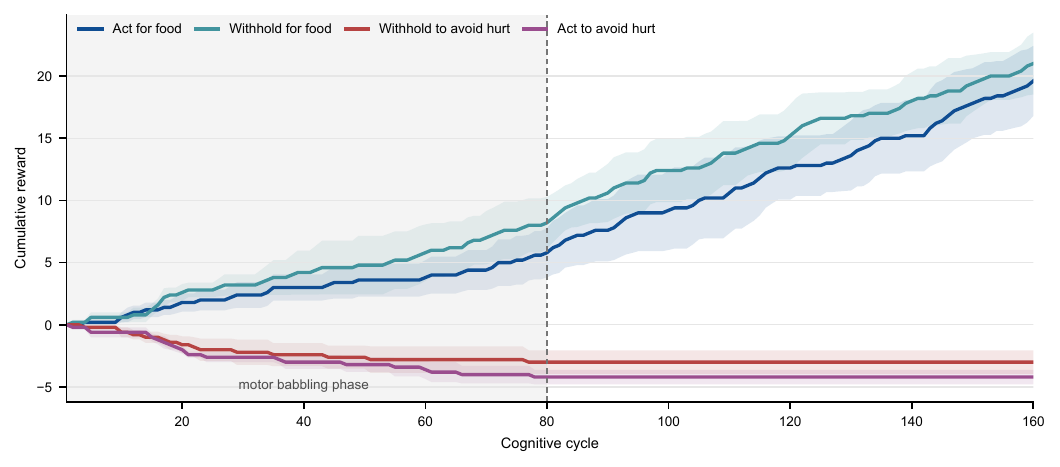}
\caption{External cumulative reward in the four light-press settings. The first
80 cognitive cycles form the motor-babbling phase; the later cycles show the
behavior selected from the derived operation-level item.}
\label{fig:reward-curve}
\end{figure}

Figure~\ref{fig:reward-curve} gives an external check on the four reasoning
results. After the motor-babbling phase, the two food cases continue to
accumulate positive reward under the behavior predicted by the theory, while the
two hurt cases stop accumulating sustained additional loss. The negative portions
of the hurt curves mainly come from exploratory mistakes during motor babbling;
once motor output is driven by the inferred operational goal, these curves become
nearly flat.

The following subsections describe the four cases in detail. For each case, the
relevant predictive schemas and motivational input are specified, then it is shown how
the proposed goal or anti-goal propagation derives the corresponding
operational motivation.

\subsection{Case 1: Do to Realize}

This case is the ordinary instrumental case. During motor babbling, the
system samples pressing under the light condition and learns that this sequence
is followed by event $food$:
\begin{equation}
(light,\op{press}) \predimply food~\truthv{1.0}{0.99}.\footnote{For simplicity of presentation, the confidence values shown here, such as $0.99$, are illustrative rather than the exact values learned in the run. The truth values used in subsequent derivation steps are treated in the same way.}
\end{equation}

Given the motivational input ``$food!~\truthv{1.0}{0.99}$'', backward goal
propagation first derives a goal for the sequence:
\begin{equation}
\begin{aligned}
&\{(light,\op{press}) \predimply food~\truthv{1.0}{0.99},~
food!~\truthv{1.0}{0.9}\} \\
&\vdash (light,\op{press})!~\truthv{1.0}{0.89}.
\end{aligned}
\end{equation}
The current observation ``$light~\truthv{1.0}{0.99}$'' then propagates this goal to
the executable operation:
\begin{equation}
\begin{aligned}
&\{(light,\op{press})!~\truthv{1.0}{0.89},~
light.~\truthv{1.0}{0.9}\} \\
&\vdash \op{press}!~\truthv{1.0}{0.8}.
\end{aligned}
\end{equation}
The derived operational goal is therefore ``$\op{press}!$'', and the system executes $\op{press}$ as the motor output. The corresponding \textit{Act for
food} curve (see Fig.~\ref{fig:reward-curve}) keeps increasing
after motor babbling, as expected when pressing realizes $food$.

\subsection{Case 2: Not Do to Avoid}

This case tests passive avoidance. During motor babbling, the system learns
that pressing under the light condition is followed by the avoided event $hurt$:
\begin{equation}
(light,\op{press}) \predimply hurt~\truthv{1.0}{0.99}.
\end{equation}
Given the motivational input ``$hurt\mbox{!`}~\truthv{1.0}{0.9}$'', backward
anti-goal propagation first derives an anti-goal for the sequence:
\begin{equation}
\begin{aligned}
&\{(light,\op{press}) \predimply hurt~\truthv{1.0}{0.99},~
hurt\mbox{!`}~\truthv{1.0}{0.9}\} \\
&\vdash (light,\op{press})\mbox{!`}~\truthv{1.0}{0.89}.
\end{aligned}
\end{equation}
The current observation ``$light~\truthv{1.0}{0.99}$'' then propagates this
anti-goal to the executable operation:
\begin{equation}
\begin{aligned}
&\{(light,\op{press})\mbox{!`}~\truthv{1.0}{0.89},~
light.~\truthv{1.0}{0.9}\} \\
&\vdash \op{press}\mbox{!`}~\truthv{1.0}{0.8}.
\end{aligned}
\end{equation}
The derived operational anti-goal is therefore ``$\op{press}\mbox{!`}$'', so
the system suppresses $\op{press}$ as a motor operation. In
Fig.~\ref{fig:reward-curve}, the \textit{Withhold to avoid hurt} curve drops
during motor babbling but becomes nearly flat afterward, matching the prediction
that pressing should be suppressed.

\subsection{Case 3: Do to Avoid}

The process of learning a preventional statement is harder than the previous cases, due to the constrastive rule in \eqref{eq:prevent-schema} require negative oberservation (\textit{i.e.}, $\neg hurt$).

The system initially observes that \textit{light} is followed by \textit{hurt} with a time interval,
\begin{equation}
(light,\_) \predimply hurt ~\truthv{1.0}{0.67},
\end{equation}
so that when event $light$ occurs, the system anticipate a future event $hurt$.
However, in motor babbling, when the system execute $\op{press}$ after the light condition, the anticipation fails. A negative observation can thus be produced,
\begin{equation}
\neg hurt ~\truthv{1.0}{0.5}.
\end{equation}
The system applies the temporal induction rule and derives ``$(light, \op{press}) \predimply \neg hurt ~\truthv{1.0}{0.31}$''. With more evidence collected, this prediction becomes stronger,
\begin{equation}\label{eq:light-press-implies-not-hurt}
(light, \op{press}) \predimply \neg hurt ~\truthv{1.0}{0.99}.
\end{equation}
Meanwhile, the object-level rule \eqref{eq:prevent-schema} enables the system to generate concepts including ``$(light, \neg\op{press})$'', ``$(light, \op{press})\imply \op{prevent}(hurt)$'' and so on, so that the system can accumulate evidence for event ``$(light, \neg\op{press})$''.\footnote{This step is indispensable because NARS cannot arbitrarily observe events that have not occurred.} With more evidence accumulated, the system learns that
\begin{equation}\label{eq:light-not-press-implies-hurt}
(light, \neg\op{press}) \predimply hurt ~\truthv{1.0}{0.99},
\end{equation}
and derives
\begin{equation}\label{eq:light-press-implies-prevent-hurt}
(light, \op{press}) \predimply \op{prevent}(hurt) ~\truthv{1.0}{0.98},
\end{equation}
by rule \eqref{eq:prevent-schema} with beliefs \eqref{eq:light-press-implies-not-hurt} and \eqref{eq:light-not-press-implies-hurt}.

The system receives the motivational input ``$hurt\mbox{!`}~\truthv{1.0}{0.9}$'', stating that $hurt$ is to be avoided. Rule~\eqref{eq:prevent-activation} then turns this anti-goal into the positive goal of preventing $hurt$:
\begin{equation}
hurt\mbox{!`}~\truthv{1.0}{0.9}
\vdash \op{prevent}(hurt)!~\truthv{1.0}{0.9} .
\end{equation}
Ordinary backward goal propagation can then use the learned prevention statement
to derive the sequence goal:
\begin{equation}
\begin{aligned}
&\{(light,\op{press}) \predimply \op{prevent}(hurt)~\truthv{1.0}{0.98},~
\op{prevent}(hurt)!~\truthv{1.0}{0.9}\} \\
&\vdash (light,\op{press})!~\truthv{1.0}{0.88} .
\end{aligned}
\end{equation}
Finally, when $light$ is observed, the sequence goal is reduced to an executable
operation goal:
\begin{equation}
\begin{aligned}
&\{(light,\op{press})!~\truthv{1.0}{0.88},~
light.~\truthv{1.0}{0.9}\} \\
&\vdash \op{press}!~\truthv{1.0}{0.79} .
\end{aligned}
\end{equation}
Thus, from the intention to avoid $hurt$, the system derives the positive
operation goal ``$\op{press}!$'' and emits $\op{press}$. This case is important
because the anti-goal does not merely suppress actions; through contrastive
inference, it can also produce a positive operation. The
\textit{Act to avoid hurt} curve (see Fig.~\ref{fig:reward-curve}) becomes nearly flat after motor babbling, because pressing prevents
further occurrences of $hurt$.

\subsection{Case 4: Not Do to Realize}

This case is the preservation counterpart of Case 3, and its learning process
is symmetric. The system initially observes that \textit{light} is followed by
\textit{food} with a time interval,
\begin{equation}
(light,\_) \predimply food~\truthv{1.0}{0.67},
\end{equation}
so that when event $light$ occurs, the system anticipates a future event $food$.
However, in motor babbling, when the system executes $\op{press}$ after the
light condition, the anticipation fails. A negative observation can thus be
produced,
\begin{equation}
\neg food~\truthv{1.0}{0.5}.
\end{equation}
The system applies the temporal induction rule and derives
``$(light,\op{press}) \predimply \neg food~\truthv{1.0}{0.31}$''. With more evidence
collected, this prediction becomes stronger,
\begin{equation}\label{eq:light-press-implies-not-food}
(light,\op{press}) \predimply \neg food~\truthv{1.0}{0.99}.
\end{equation}
Meanwhile, the object-level rule \eqref{eq:prevent-schema} enables the system
to accumulate evidence for the contrastive event ``$(light,\neg\op{press})$''.
With more evidence accumulated, the system learns that
\begin{equation}\label{eq:light-not-press-implies-food}
(light,\neg\op{press}) \predimply food~\truthv{1.0}{0.99},
\end{equation}
and derives
\begin{equation}\label{eq:light-press-implies-prevent-food}
(light,\op{press}) \predimply \op{prevent}(food)~\truthv{1.0}{0.98},
\end{equation}
by rule \eqref{eq:prevent-schema} with beliefs
\eqref{eq:light-press-implies-not-food} and
\eqref{eq:light-not-press-implies-food}.

The system receives the motivational input ``$food!~\truthv{1.0}{0.9}$'',
stating that $food$ is to be realized. Rule~\eqref{eq:prevent-activation} then
turns this goal into the anti-goal of preventing $food$:
\begin{equation}
food!~\truthv{1.0}{0.9}
\vdash \op{prevent}(food)\mbox{!`}~\truthv{1.0}{0.9} .
\end{equation}
Backward anti-goal propagation can then use the learned prevention statement to
derive the sequence anti-goal:
\begin{equation}
\begin{aligned}
&\{(light,\op{press}) \predimply \op{prevent}(food)~\truthv{1.0}{0.98},~
\op{prevent}(food)\mbox{!`}~\truthv{1.0}{0.9}\} \\
&\vdash (light,\op{press})\mbox{!`}~\truthv{1.0}{0.88} .
\end{aligned}
\end{equation}
Finally, when $light$ is observed, the sequence anti-goal is reduced to an
executable operation anti-goal:
\begin{equation}
\begin{aligned}
&\{(light,\op{press})\mbox{!`}~\truthv{1.0}{0.88},~
light.~\truthv{1.0}{0.9}\} \\
&\vdash \op{press}\mbox{!`}~\truthv{1.0}{0.79} .
\end{aligned}
\end{equation}
Thus, from the intention to realize $food$, the system derives the operational
anti-goal ``$\op{press}\mbox{!`}$'' and emits no motor operation. In this case,
the system suppresses an action because that action prevents a desired event.
The \textit{Withhold for food} curve (see Fig.~\ref{fig:reward-curve}) continues to increase after motor babbling, confirming that suppressing the operation preserves the desired event
in this setting.

\section{Discussion}
\paragraph{On Theoretical Assumptions and Scope.}
 In IL, the idealized situation, the framework assumes a clear contrast between
what the system treats as to-be-realized and what it treats as to-be-avoided.
This contrast provides a reference point for defining goals and anti-goals, but
it should not be interpreted as saying that the system must always classify
every event motivationally in NAL:
Once the framework is placed under AIKR, motivation becomes incomplete and
revisable. An event may be neither desired nor avoided because the system has no
relevant motivational evidence about it. Such an event should remain
motivationally \textit{unknown}, rather than being forced into either side of
the idealized contrast. This design is also what makes the extension
applicable to \textit{open world}~\cite{xu2024ow}: motivations need not be fully prespecified, but can be introduced, strengthened, weakened, or revised as new experience becomes
available.

In general, the motivations of an adaptive system working with insufficient knolwedge and resoures may come from bodily presets, environmental inputs such as linguistic instructions, and internal derivation; these motivations may event conflict with one another; over time, these sources may lead to the system's own distinctive purposes.~\cite{wang2012motivation} 
To keep the present paper focused, the acquisition of motivation is not modeled in detail. The aim here is only to extend NAL's goal-reasoning framework with anti-goals and prevention; a full theory of motivational learning remains an important topic for future work.

\paragraph{On Asymmetry between the Cases.}
A related behavioral finding suggests that these cases are not symmetric. Guitart-Masip et al.~\cite{guitart-masip2012go-nogo} studied a task in which human participants had to learn when to act and when to withhold action under reward and punishment. They found that people more readily learned to act for reward and to withhold action to avoid punishment, whereas acting to avoid punishment and withholding action to obtain reward were learned less readily.
This behavioral asymmetry is consistent with the theoretical construction proposed here: Case 1 and Case 2 are direct goal or anti-goal propagation, whereas Case 3 and Case 4 additionally require learning a prevention relation from negative observation and contrastive evidence. Thus, the behavioral result supports the distinction made by the present framework, even though the present theory and the account from \cite{guitart-masip2012go-nogo} explain the behavioral asymmetry by very different mechanisms.

\paragraph{On the Difference from Reinforcement Learning.}
The proposed distinction between goals and anti-goals is also different from the usual treatment of reward in reinforcement learning (RL)~\cite{sutton2018rl-intro}. In a standard reinforcement-learning formulation, the agent selects an action at each decision step by comparing the expected returns of available actions, possibly including a no-operation action if it is included in the action set. Under this view, a negative reward does not by itself mean that the corresponding state or event is an anti-goal in the present sense. If all available actions have negative expected returns, the action with a negative expected return may still be selected, just because it has the highest expected return. Therefore, the sign of a reward value does not directly encode the distinction between desired and avoided events. By contrast, in NAL extended herein, an operation is not selected merely because it is better than the other available operations. An operation is executed only when it receives sufficient motivational support; otherwise, the system may simply remain silent and emit no operation. In this sense, no-operation is not necessarily just another competing action, but can be the default condition of the system. These different theoretical presuppositions may lead to different practical consequences. A systematic comparison between the RL and NAL is beyond the scope of this paper, but it is an important topic for further research.

\section{Conclusion}

By analyzing a paradoxical example, this paper exposes a theoretical ambiguity
in previous NAL accounts of motivational reasoning. The proposed response is not
to replace the existing theory of goals, but to complete it with a minimal
extension to NAL.

The theoretical contribution of this paper is threefold. First, it gives
avoidance its own evidential status by separating anti-goals from ordinary goals
whose event content happens to be negated. This separation removes the ambiguity
in goal negation and makes clear why pursuing the absence of an event is not the
same as avoiding the event itself.

Second, the paper gives a general principle for choosing desire-value functions
in backward motivational inference. The principle is to justify a backward rule
by relating its motivational reading to the corresponding forward inference
rule.

Third, the paper introduces active prevention as a bridge between anti-goal reasoning and ordinary goal reasoning. Active prevention is not reduced to the mere observation that an event is absent after an action; it requires contrastive evidence that the action changes what would otherwise be expected.

\subsubsection*{Acknowledgments.} The author appreciates the advice from Dr. Pei Wang; discussions and debates with him were crucial during the development of this idea and theory. The proposed solution could not have been developed without his guidance and inspiration.
This paper also benefits from discussions with Boyang Xu and Bojie Feng.

\bibliographystyle{splncs04}
\bibliography{reference}

\end{document}